\documentclass[useAMS,usenatbib]{mn2e}
\usepackage{graphicx}
\usepackage{txfonts}
\usepackage{natbib}
\usepackage{multicol}
\usepackage{graphics} 
\usepackage[T1]{fontenc}
\newcommand\etal{et~al.~}

\newcommand\pegase{\textsc{P\'egase}}

\title[Old stars in starbursting z=3.8 radiogalaxies ]{Starburst and
old stellar population in the z=3.8 radiogalaxies 4C~41.17 and TN~J2007$-$1316}

\author[B.Rocca-Volmerange et al.]
{B. Rocca-Volmerange$^{1,4}$\thanks{E-mail:
rocca@iap.fr (BRV)}, G.~Drouart$^{2,1}$\footnotemark[1]\thanks{ESO Doctoral Fellow}, C.~De Breuck$^{2}$, J.~Vernet$^{2}$,
N.~Seymour$^{3}$,
 \newauthor
D.~Wylezalek$^{2}$, M. Lehnert$^{1}$, N.Nesvadba$^{5}$  and M. Fioc$^{1}$\\
$^{1}$Institut d'Astrophysique de Paris, Universit\'e Pierre et Marie Curie/CNRS,98bis Bd Arago, F-75014 Paris, France\\
$^{2}$European Southern Observatory, Karl-Schwarzschild Strasse, 85748 Garching bei M\"{u}nchen, Germany\\
$^{3}$AE (CSIRO Astronomy \& Space Science), P.O. Box 76, Epping, NSW 1710, Australia\\
$^{4}$Universit\'e Paris-SUD, 91405, Orsay Cedex, France\\
$^{5}$Institut d' Astrophysique Spatiale, CNRS, Universit\'e Paris-Sud, 91405, Orsay, France\\
}

\date{Received 2012 August 10; accepted 2012 November 5}

\def\LaTeX{L\kern-.36em\raise.3ex\hbox{a}\kern-.15em
  T\kern-.1667em\lower.7ex\hbox{E}\kern-.125emX}

\begin{document}

\maketitle

\label{firstpage}

\begin{abstract} 
Using the new evolutionary code \pegase.3, we made an
evolutionary  spectral  synthesis  of  the  optical-IR-submm  spectral
energy distribution of two  distant ($z=3.8$) radio galaxies, 4C~41.17
and TN~J2007$-$1316.  These two  radio galaxies were selected from the
HeRG\'E (Herschel Radio Galaxies  Evolution) Project in particular for
their faint AGN contribution  and because they show evidence of a
large stellar contribution to their bolometric luminosity.  \pegase.3
coherently models  the reprocessing of the stellar  luminosity to dust
emission   allowing  us  to   build  UV-to-IR-submm   spectral  energy
distribution libraries which  then can be used to  fit spectral energy
distributions in  the observer's  frame.  Our principal  conclusion is
that a single  stellar population is insufficient to  fit the spectral
energy distributions of either radio  galaxy.  Our best fits are a sum
of two evolving stellar populations  -- a recent starburst plus an old
population  --  plus the  thermal  emission  from  an active  galactic
nucleus (which provides a good  fit to the mid-IR emission). The
two stellar components are: i) a massive ($\simeq 10^{11}\,$M$\odot$)
starburst   $\simeq$30\,   Myrs  after formation, which   is  required
simultaneously to fit the far-IR  {\it Herschel} to submm data and the
optical data  ii) an  older  massive ($\simeq  10^{11-12}$\,M$\odot$)
early-type  galaxy  population,  $\simeq$\  1.0\,Gyr  old,  which  is
required  principally to fit  the mid-IR  {\it Spitzer}/IRAC  data.  A
young population  alone is insufficient because  an evolved giant-star
population produces  a 1$\,\mu$m rest-frame peak which  is observed in
the IRAC photometry.  This discovery confirms that much of the stellar
populations  in high  redshift radio  galaxies are  formed  by massive
starbursts  in  the  early  universe. Gas-rich  mergers  and/or
jet-cloud  interactions are  favored for  triggering the  intense star
formation necessary to explain the properties of their spectral energy
distributions.  The discovery of similar  characteristics in two distant radio
galaxies suggests  that multiple stellar  populations, one old  and one
young, may be  a generic feature of the luminous infrared radio
galaxy population as a whole.
\end{abstract}


\section{\bf Introduction}
Large scale cosmological simulations offer the prospect of constraining
theories of galaxy formation and probing the appropriateness of
cosmological parameters.  However, hydrodynamical and N$-$body simulations
meet various difficulties in fitting observations.  The reasons
for these difficulties is that there are many physical processes and
timescales which are not well understood such as: the appropriate initial
conditions for the models, the detailed structure of dark matter halos,
the timescale over which mass is accumulated into the halo and onto the
galaxy,  how stars form in detail, the relative role of major and minor
mergers and their importance over cosmic time, the effect of the density of
the galaxian environment and the impact of stellar and active galactic
nuclei (AGN) feedback.  Radio galaxies are known to be massive elliptical
types which formed in the early Universe \citep{vbg98,pen01,zi03,lacy2011}. 
The puzzling
Hubble $K-z$\ diagram, which is bounded on the bright side principally
by powerful radio galaxies  \citep{ll84,deb02a}, allowed us 
to identify in the comoving rest$-$frame the population of
elliptical galaxies. This population is young but already  massive (10$^{12}$M$_{\odot}$)
out to high $z=$4 redshifts \citep{rv04}.
On the other hand the mid$-$IR, far$-$IR and sub$-$millimeter emission from
distant radio galaxies were interpreted as starbursts mainly initiated
by major mergers \citep{deb10,ivi08,ivi12,eng10,sey10,sey12}. These dichotomous results:
the old population from the Hubble $K-z$\ diagram and the young population 
from the need for high rates of star formation
in radio galaxies imply that we are required to simultaneously follow
the passive evolution of the galaxy as well as that of the on$-$going
starburst to explain the overall spectral energy distribution (SED). To
do this simultaneously requires a refined spectrophotometric model
such as our new version of \pegase.  The  code \pegase.3 offers the
possibility to interpret observations in the observer's frame by using
UV-to-submm synthetic libraries built for a wide range of ages. The
template SEDs take into account the local z=0  templates corrected
for evolution of the stellar population (e$-$correction) and redshift
(k$-$correction).  In the present paper we tackle a precise, heretofore
still  debated key  issue-namely whether the most distant  galaxies,
including  radio galaxies, are pure starbursts  initiated by on$-$going
mergers or are forming stars continuously on timescales longer than
that generally appropriate for mergers.  In  radio galaxies, there is
also the additional difficulty of the thermal emission from the AGN
(perhaps a torus) and of the scattered light from the AGN which both 
contribute to the
overall spectral energy distribution, especially in the mid$-$IR and UV
respectively \citep{dro12}.
We chose two radio galaxies in particular, 4C~41.17 and  TN~J2007$-$1316,
both at z=3.8, because of their small AGN contribution
to the overall SED.  We apply the following two selection criteria:
(i)  relatively faint scattered  light contribution  in  the
rest-frame UV as determined from polarimetric measurements. Polarimetric
measurements are the only way to ensure that the continuum is mostly
of stellar origin. Indeed those two galaxies are the only cases where
a clear signature of photospheric absorption lines has been detected
\citep{dey97} (ii) faint contribution of the AGN in the 
rest-frame 8-12$\mu$m range of the SED.  To fit the SED, one and two
stellar components are successively tried from an automatic procedure
which minimizes the $\chi^2$.  We focus on the main properties (type,
mass, age and star-formation history) of the two stellar populations.
In \S~2, we present the observations of the two radio galaxies for
which we have built the continuous optical-IR-submm SEDs after correcting
for instrumental effects (differing apertures, flux calibrations, etc.).
\pegase.3 and the fitting method are described in \S~3.  \S~4 
presents the best fit results of SEDs in the  observer's frame 
obtained using two evolved stellar  populations plus  a  simple
AGN  model.  Finally, we provide some thoughts on the implications of
these results in \S~5 and our conclusions in the last section.  The adopted
cosmological parameters are H$_0=70km.s^{-1}Mpc^{-1}, \Omega_M=0.3,
\Omega_{\Lambda}=0.7$.

\section {\bf Observations}

The two radio galaxies 4C~41.17 ($z\simeq 3.80$) and TN~J2007$-$1316
($z\simeq 3.84$) are part of the sample of ultra-steep spectrum radio
sources (USS; $\alpha<-1.3$; $S_{\nu} \propto \nu^{\alpha}$) designed
specifically to increase the probability of discovering distant,
$z>3$, radio galaxies \citep{deb00}.  The two targets
were specifically selected because of their relatively low level
of AGN activity in the rest-frame UV/optical and their strong
photospheric and dust signatures respectively seen in their
optical continuum emission and in the cold dust-grain dominated
far-IR emission.  The {\it Spitzer} High Redshift Radio Galaxies (SHzRG)
sample, which these two galaxies are also part of, were augmented with
$K-$band photometry \citep{sey07,sey08}.  Most of the complementary
optical and submm photometry and data reduction procedures have been
described previously \citep{deb10}.  In addition, both
galaxies have recently been observed by {\it Herschel} with the PACS (100
and 160$\,\mu$m) and SPIRE (250, 350 and 500$\,\mu$m) instruments, 
see also for 4C~41.17 \citep{wy12}, and are part of the HeRG\'E (Herschel
Radio Galaxy Evolution) Project \citep{sey12}.  Optical to submm
spectral energy distributions, sampling the rest-frame photospheric
(UV/optical/near-IR) stellar emission to emission from cold grains (far-IR) are carefully
built paying specially attention to aperture effects, flux calibration,
and flux transmission of the filters used for the observations and
removing the contribution to the fluxes from strong emission lines. For
the two galaxies the synchrotron emission is considered as negligible in
the wavelength domain hereafter analyzed due to the steep radio spectra
of the cores.

\subsection{\bf The radio galaxy 4C41.17}
The radio galaxy 4C~41.17 ($z=3.800\pm0.003$) is the archetype of distant radio galaxies
\citep{ch90}.  A detailed multi-frequency
radio analysis with the VLA and MERLIN shows that the radio structure
is associated with high resolution optical imaging {\it HST} data
\citep{car94,mi92}. The galaxy is detected
with a good signal-to-noise ratio and is spatially resolved at the $0.1''$
{\it HST} resolution. Based on both the C$_{III]}$\ to C$_{IV}$\ line
ratio and the strength of the C$_{IV}$\ line, \citep{bi00}
suggested that there is an interaction of the high-powered jet with a
dense cloud in the halo of 4C~41.17.  Such an interaction through high
speed shocks ($\sim1000\,$km/s), leads to strong compression of some
of the gas, inducing star formation.  The bright, spatially extended
rest-frame UV continuum emission from this galaxy, aligned with the
radio axis, is unpolarised ($P_{2\sigma} < 2.4\%$) and shows tell-tale
stellar absorption features indicative of on-going star formation
\citep{dey97}.  Two massive ($M_{dyn}= 6\times 10^{10}\,M_\odot$)
components have been identified from CO(J=4$-$3) observations with the IRAM
interferometer and interpreted as an evidence of mergers \citep{deb05}.  
However, the CO (J=1$-$0) emission line is not detected and
the lower limits obtained are typical of molecular gas in starbursts
\citep{pap05}.  To complete the broad  wavelength coverage
of the SED of 4C~41.17, we use the total of the flux from 4C~41.17
within 6 arcsec field in the $K_s-$band, which is basically free of
emission from strong optical emission lines \citep{gr94}. 
For the submillimeter emission we use the results from the JCMT
\citep{du94} and IRAM \citep{chk94}.

The  velocity and dispersion fields were mapped with the integral
field spectrograph TIGER/CFHT \citep{ad97} with a 0.61 arcsec
spatial sampling. Radial velocities are essentially negative as
the velocity field shows bow shocks with high velocity dispersions.
Perpendicular to the main radio axis, a South-West extension of velocity
$=-600\,$kms$^{-1}$\ and velocity dispersion $= 2000\,$kms$^{-1}$\ is 
visible.  Table~\ref{data4C41.17} contains the photometry data we
used to construct the SED of 4C~41.17. The upper limits and data not
used by the fitting procedure are also provided for completeness and are
also show in subsequent figures.

\begin{table}
\caption[]{Photometric data, in red on Fig.~\ref{SaSB4C4117} and
Fig.~\ref{SEDs4C41.17}, of the radio galaxy 4C41.17 including 
new {\it Herschel}/SPIRE and PACS observations 
($\alpha(\rm J2000)=06^h 50^m52.098^s$, 
$\delta(\rm J2000)=+41\degr 30\arcmin 30.53\arcsec$).
-tot- in column 4 means that galaxy photometry is totally 
measured within the instrument aperture.
Complementary data, lower rows, in green on Fig.~\ref{SaSB4C4117}
and Fig.~\ref{SEDs4C41.17} are not used for fits, but are provided
for completeness.}
\footnotesize
\begin{tabular}{|c|c|c|c|c|} 
\hline
Filter($\lambda_c\mu$m) & FWMH& F$_\nu \pm\Delta_\nu$($\mu$Jy) 
  & Apert.& Refer. \\
\hline 
$HST\_$F702W(0.7) &0.15 & 5.0 $\pm$  0.4  &$5.0''^2$& $a$\\
$KPNO_I$(0.9)  &0.22 & 4.5  $\pm$  2.6  & $15''$  & $b$  \\
$NIRC_J$(1.25)  &0.29 & 5.6  $\pm$  1.1  & $2''$  & $c$  \\
$NIRC_{Ks}$(2.15) &0.33 & 13.6 $\pm$  2.8  &  $8.0''$  & $j$ \\
IRAC1(3.6)  &0.74 & 23.4 $\pm$  2.4  &  $12''$  & $d$  \\
IRAC2(4.5)  & 1.0 & 27.5 $\pm$  2.8  &  $12''$  & $d$  \\
IRAC3(5.8)  &1.4  &35.6  $\pm$  3.7  &  $12''$  & $d$  \\
IRAC4(8.0)  &2.8  &36.5  $\pm$  3.5  &  $12''$  & $d$  \\
PACS(170.)  &80  & 16.2 $\pm$  6. 7 (+3) &  tot  & $e$  \\
SPIRE(250.)  &100  & 35.8 $\pm$ 3.5  (+3) &  tot  &  $e$  \\
SPIRE(350.)  &150  & 43.1 $\pm$ 3.7  (+3) &  tot  &  $e$  \\
SPIRE(500.)  &200  & 38.0 $\pm$  4.5  (+3)&  tot  &  $e$  \\
UKT14(800.)  &--  &17.4  $\pm$  3.1  (+3) &  tot  &  $h$  \\
SCUBA(850.)  &--  &12.1  $\pm$  0. 9  (+3)&  tot  &  $g$  \\
IRAM(1200.)  &--  &4.4  $\pm$  0.4  (+3) &  tot  &  $f$  \\
IRAM(1300.)  &--  &2.5  $\pm$  0.4  (+3) &  tot  &  $i$  \\
 \hline
IRS(16)  &--  & upper limit  & tot  & $d$    \\
MIPS(24)   &6.6  & 3.7  $\pm$0.4 (+2)  &  tot  & $d$  \\
PACS(70)   &35  &  upper limit  &  tot  & $e$  \\
\hline
\end{tabular} 
\ \\
$a$:\citep{mi92} \ \ \ \ \ \ \ \ \ \ \ \ \ \ \ \ \ \ $f$:\citep{gre07}\\
$b$:\citep{ch90} \ \ \ \ \ \ \ \ \ \ \ \ \ \ \ \ $g$:\citep{arc01}\\
$c$:\citep{vbg98}\ \ \ \ \ \ \ \ \ \ \ \ \ \ \  $h$:\citep{du94} \\
$d$:\citep{deb10} \ \ \ \  \ \  $i$:\citep{chk94}\\
$e$:this paper , see also \citep{wy12}\\
$j$:\citep{gr94}\\
\label{data4C41.17}
\end{table}

\subsection{\bf The radio galaxy TN~J2007$-$1316, z=3.84 }

\begin{figure}
\centering
\includegraphics[angle=0,width=8cm]{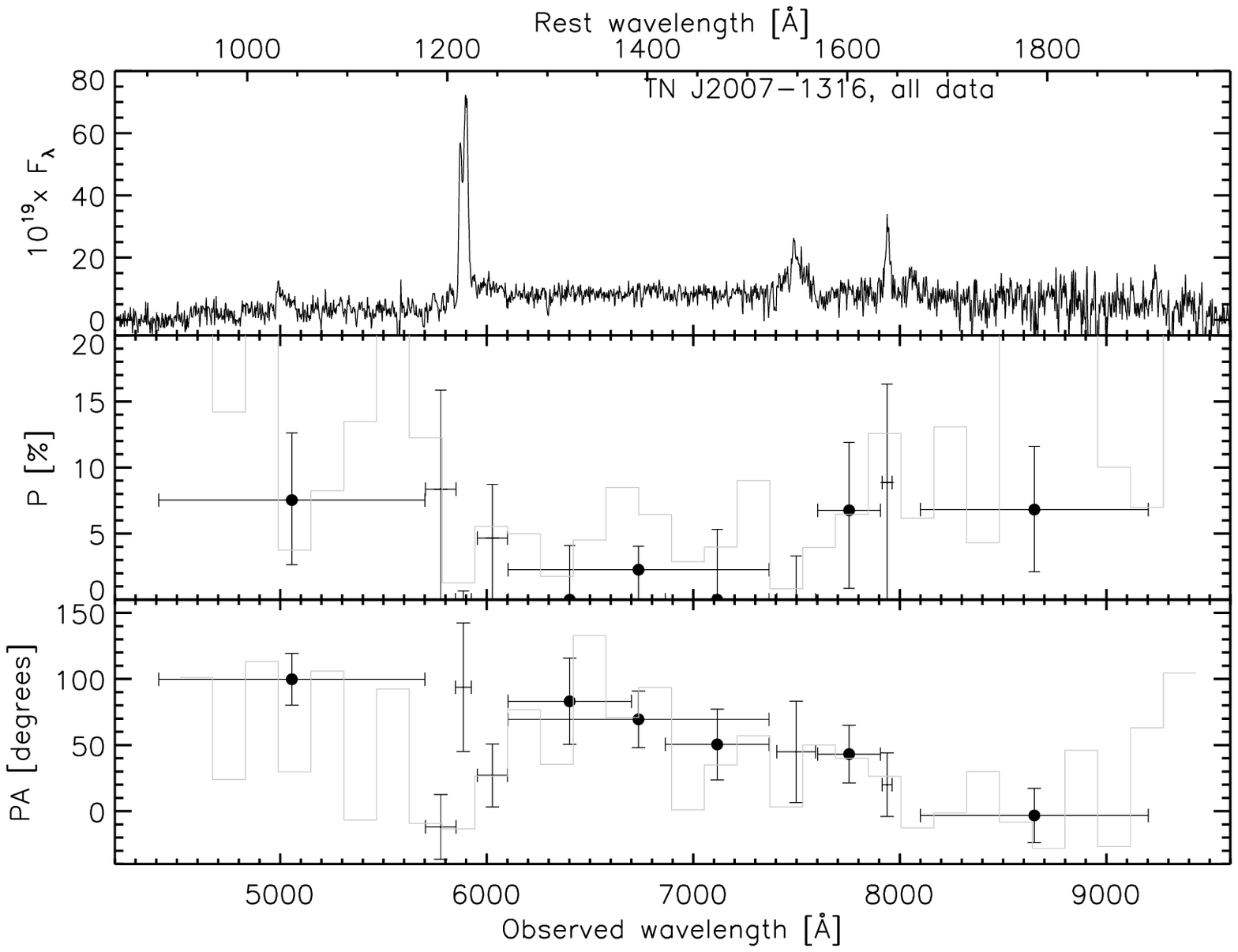}
\includegraphics[angle=+90,width=8cm]{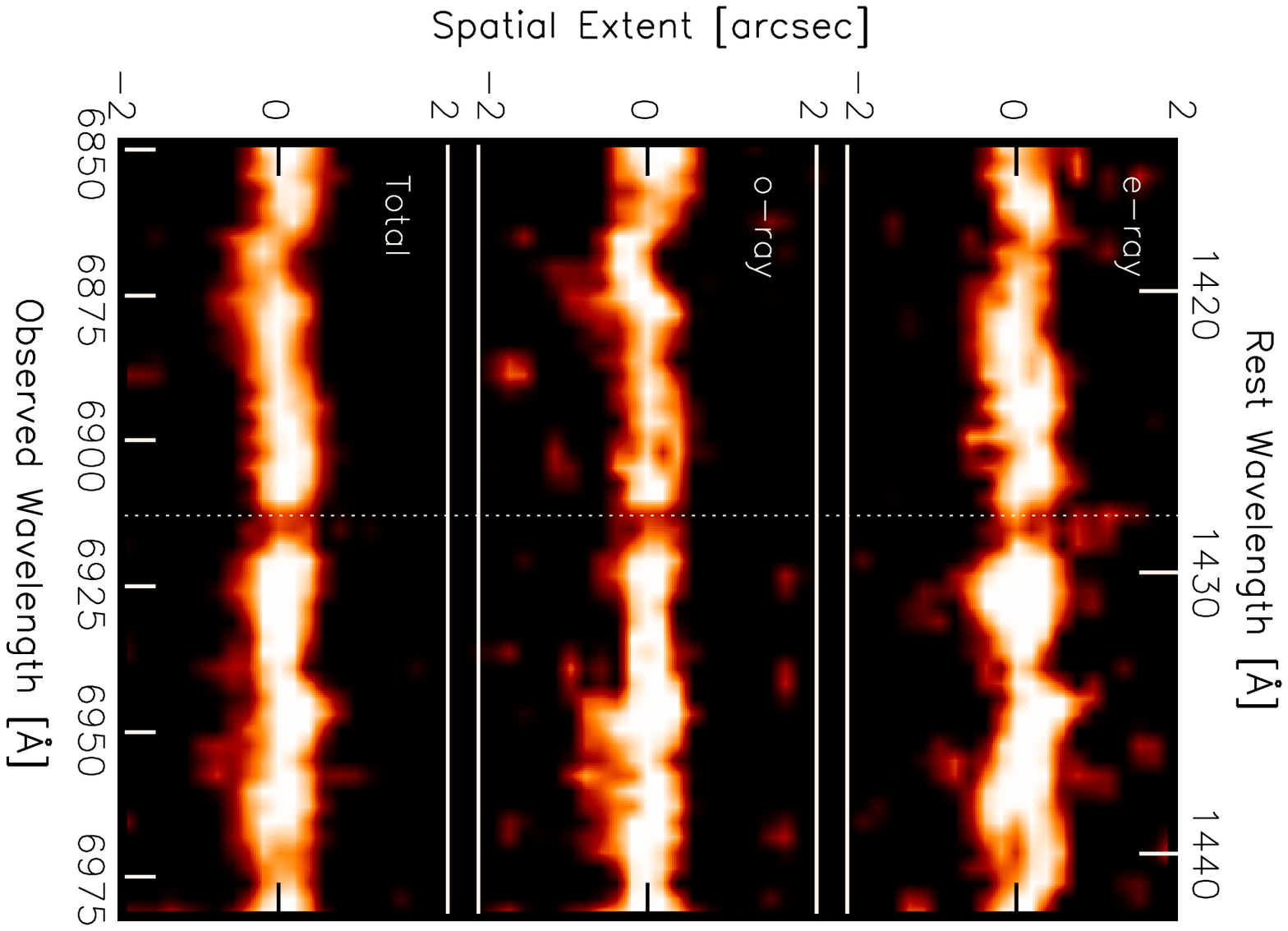}
\caption{VLT spectropolarimetry of TN~J2007$-$1316. the top panel is
  divided in three parts.  The upper shows the total intensity
  spectrum, the middle plot is the percentage polarization, and the
  bottom panel the polarization angle, as function of observed
  wavelengths. Continuum bins are denoted by a dot. Vertical error
  bars denote $1\,\sigma$\ uncertainties. The grey histograms show the
  data with a $150\,$ \AA\ binning.  The bottom panel is the 2D
  spectrogram, showing evidences of stellar signatures.}
\label{pol_tn2007}
\end{figure}

TN~J2007$-$1316 (or WN~J2007$-$1316)is also an ultra steep spectrum
radio  source selected from the 365\,MHz TEXAS  survey \citep{dou96} and 
the NVSS survey \citep{con98}.  It was also detected in the 352\,MHz WISH 
survey and is part of the USS sample \citep{deb02b}.  
Spectropolarimetry  of  TN~J2007$-$1316 obtained with FORS1 on the VLT on 
3 nights from 2002~May~6 to 8(Fig.~\ref{pol_tn2007}).  
For these observations, the
300V grism was used with a $1''$\ wide slit oriented North-South and had a
resolution of $\sim10\,$ \AA\ (FWHM).  Conditions  were photometric  with
seeing between  0.5-0.9 arcsecs.  During each  night, four  equal-length
exposures were taken using four different orientations of the half-wave
plate  (0\degr, 45\degr, 22\fdg5,  67\fdg5).  The  total integration time
was 27600s.  The initial  data reduction followed standard procedures
in the NOAO IRAF package. Because the  object is relatively faint in
the single 35 or 40  minute exposures (each split into o- and e-rays),
we needed to pay particular attention to extracting the same physical
apertures in each of the individual spectra. We therefore first contructed
the total  2-dimensional o- and e-ray spectra, and defined 1$\times$2.4
arcsec$^2$ wide apertures from these spectra.  We then used the aperture
and trace of these spectra to extract the 12 o-ray and 12 e-ray spectra,
using  the  same  linear dispersion ($2.645\,$ \AA\ /pix) in order to
calculate the polarization in equally-sized spectral bins. Finally, we
combined these spectra using the median to construct the eight spectra
needed to calculate the polarization vector. We  followed the procedures
of \citep{ver01} to calculate  the  polarization percentage and
position angle. We checked the polarisation angle offset between
the half-wave plate coordinate  and  sky coordinates against  values
obtained  for  the polarized standard stars Vela1 and Hiltner 652;
our values are within $<$1\degr\ from the published values, and the
polarization percentage within $0.1\,\%$. Table~\ref{dataTNJ2007}
provides the results and data source references.

\begin{table}
\caption[]{Photometric data of the radio galaxy TN~J2007$-$1316 (in
red on Fig.~\ref{SEDsTNJ2007}), $\alpha(\rm J2000)=20^h 07^m 53^s.23$,
$\delta(J2000)=-13\degr 16\arcmin 43.6\arcsec$.  Complementary data
(bottom lines) are not used for fits, only given for completeness.
The continuum polarization percentage accurate to within $0.1\,\%$.}
\footnotesize
\begin{tabular}{|c|c|c|c|c|} \hline
Filter($\lambda_c(\mu$m)) & FWMH& F$_\nu \pm\Delta_\nu  (\mu $Jy) & Apert.& Refer. \\
\hline 
R(0.65)  &  0.06  &2.1 $\pm$ 0.4  &2.0$"^2$ & $a$\\
CFHT$_I$(0.9)& 0.22 & 2.6 $\pm$ 0.15 &2.0$"^2$ & $a$\\
ISAAC$_H$(1.65)& 0.3& 9.6 $\pm$ 1.0 & tot & $a$\\
UKIRT$_K$(2.2)& 0.3& 28.4 $\pm$ 1.9 &2.0$"^2$ & $d$\\
IRAC1(3.6)  &  0.74  &46.6  $\pm$  4.8  &tot  & $a$ \\
IRAC2(4.5)  &  1.0  &52.7  $\pm$  5.7  &  tot  & $a$ \\
SPIRE(250.)  &  100.  &13.8  $\pm$ 6.1 (+3)  &  tot  & $a$  \\
SPIRE(350.)  &  150.  & 16.5$\pm$ 6.4 (+3)&  tot  & $a$ \\
SPIRE(500.)  &  200. & 7.6 $\pm$  3.3 (+3) &  tot  & $a$ \\
SCUBA(850.) &  --  &  5.8 $\pm$  1.5(+3)&  tot & $b$ \\
 \hline
IRAC3(5.8)  &  1.4  & upper limit  &  tot  & $a$  \\
IRAC4(8.0)  &  2.8  &135.1  $\pm$  16.9  &  tot  & $a$ \\
IRS (16.0)  &  6.0  &378.0  $\pm$ 113.  &  tot  & $c$ \\
MIPS1 (24.)  &  6.6  & 385.0 $\pm$  40.00 &  tot  & $c$\\
PACS(105)  &  40.  & upper limit &  tot  & $a$ \\
PACS(170)  &  80.  & upper limit &  tot  & $a$ \\
\hline
\end{tabular}
\\
$a$: this paper\ \ \ \ \ \ \ \ \ \ \ \ \ \ \ \ \ \ \ \ \\
$b$: \citep{reu04} \\
$c$: \citep{deb10} \ \ \ \ \ \ \ \ \\
$d$: \citep{bor07}\\
\label{dataTNJ2007}
\end{table}

\section{The Evolutionary  Code  \pegase.3 }  

\begin{figure}
\centering
\includegraphics[angle=0,width=9cm]{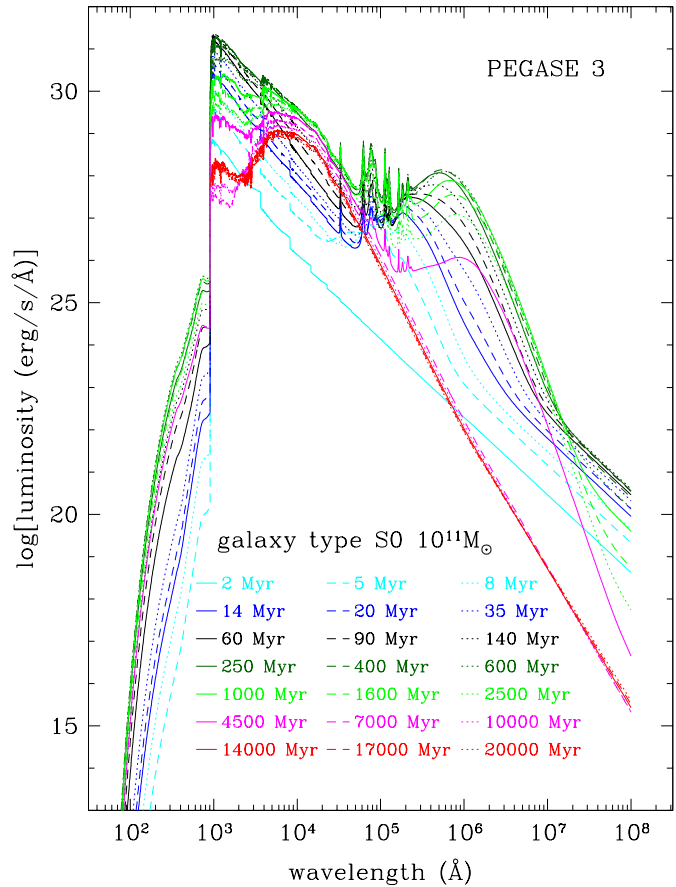}
\caption{An example of SED templates for the $10^{11}\,M_\odot$ S0-type
galaxy at various stages of evolution as computed with \pegase.3. The
$1\,\mu$m peak of giant stars appears at ages $\geq$ 1Gyr. Input
parameters are: star formation rate = $2. 10^{-3}\times$ current gas
mass per Myr, infall time scale of 100Myrs, galactic winds eject gas
and dust at 5 Gyr, a Kroupa IMF, and spheroidal extinction model (see
text for details).}
\label{S0spectra}
\end{figure}

The new version of \pegase.3 
(Fioc, Rocca-Volmerange and Dwek, in preparation) 
predicts
the evolution of the stellar continuum, of the the metal and dust content,
the consequent attenuation and the re-emission of the UV/optical continuum by
dust in a self-consistent manner. With this new version \pegase.3, updated from 
\pegase.2 (www.iap.fr\slash pegase,\citep{fi97,fi99b}), synthetic
spectral energy distributions are continuously created from the far-UV
to the sub-millimeter wavelengths at all ages (www.iap.fr\slash pegase,
\citep{fi97}.  We built  evolutionary scenarios which
are consistent with the observed SEDs (and colors) of the full range of
Hubble types at z=0.  For convenience, we refer to these SEDs by their
best fit of morphological Hubble types at z=0, with the understanding that
they, by definition, represent the SEDs of galaxy representative their
respective Hubble types in the local Universe.
In \pegase.3, the chemical enrichment history is computed in an one$-$zone
model, following the evolution of helium and most importantly, metals such
as O, Ne, Mg, Si, S, Ca, C, Fe  and N.  We used the yields 
from \citep{ma01} for  low$-$\ and  intermediate$-$mass  stars; 
those  of \citep{ww95} and \citep{po98} for high$-$mass stars;
and model W7 of \citep{th86} for type Ia supernovae, see
details in Fioc et al. in preparation.  We assumed that dust formed
in circumstellar  environments \citep{dw98}:  in the stellar  winds of
AGB  and Wolf$-$Rayet stars,  only carbon  dust is  produced if the C/O
ratio  is  larger than  $1$\  and  only  silicate dust  otherwise;  in
supernovae, as CO is unlikely to form, both carbonaceous and silicate
grains may form but with a lower  efficiency than in other sources
of dust. We compute  the dust emission by distinguishing between two
components: star-forming HII regions and the diffuse interstellar
medium.  Monte Carlo simulations of radiative transfer  are  computed
including  scattering  by  dust  grains,  for geometries typical  of disks
(exponential profiles for  both stars and dust) and elliptical galaxies
(King profiles).  The path lengths of photons before escape, absorption,
or scattering are obtained from the method of virtual interactions \citep{va99}. 
 The attenuation is  calculated for a grid
of optical depths, albedos and asymmetry  parameters, and the value at
each wavelength is then interpolated on this grid. For each grain
type and size, the probability distribution of temperatures is derived
following  the procedure detailed in \citep{gd89} 
i.e. with stochastic heating but assuming continuous cooling of grains.

To  model  a  medium  of  higher  gas  density  than  the  classical
interstellar  medium  ISM  we  consider the  optical  depth  depends
simultaneously on  the column density NHI and the current metallicity
\citep{gu87}. Following the formalism of
\citep{dw98}  based  on  the  respective fractions due to scattering
$\kappa_{sca}$ and to extinction $\kappa_{\rm  abs}$, the effective
optical depth is:

 $$\tau_{\rm eff}= (\kappa_{\rm abs} + \kappa_{\rm sca}) \times  K\,.\,NHI_{\rm ISM}. (Z_C+Z_{\rm Si}). m_{\rm H} . 1.4$$

In case of denser absorbing/scattering media, the column density
is multiplied by a $K$ factor  where 
$NHI_{\rm ISM}= 6.8\times10^{21}\,$atoms\,cm$^{-2}$\ measured by mass unit of carbon and silicon
in the interstellar medium of our Galaxy.

\subsection{\bf \pegase.3 libraries by types}
Fig.~\ref{S0spectra} shows the example of the synthetic 
templates for a S0$-$type galaxy at various ages as computed  by  the  code \pegase.3. 
Various synthetic libraries of SEDs evolving from 
ages of 0 to 20 Gyr are constructed to simulate the evolution of 
instantaneous starbursts and of galaxy types of Hubble sequence. 
Each star formation scenario is
defined predominantly by a set of four parameters: star formation law,
infall timescale,the epoch of galactic winds for early  types and stellar
initial mass function (see \pegase.2 readme on www.iap.fr/pegase). The
parameter  set  of each  type  \citep{leb02,fi99a} 
is chosen to predict $z=0$\ SEDs and colors
comparable by types to  local observations \citep{fi97}. 
The
robustness of  the adopted scenarios is checked on distant galaxies
from the UV to the near$-$infrared $K-$band \citep{rv04}. They are also tested 
on faint  galaxy surveys analyzed by spectral types in
the mid$-$IR \citep{rv07} and the UV$-$optical$-$nearIR domains \citep{fi99a}. 
Moreover  they are used for photometric redshift prediction out
high redshift ($z$=4) which shows substantial agreement with redshifts
obtained spectroscopically \citep{leb02}.
Instantaneous starbursts are defined such that they form their stars
over a period of 1~Myr.  Then they passively evolve to the phase where
their far$-$IR emission is dominated by the circumstellar emission of
AGB stars and SN (the post-burst phase). The code computes by transfer this IR
emission based on the efficiency at which the UV and optical photons
are absorbed. Besides the SEDs by types, mass and age of the stellar
populations are further  outputs from the fitting procedure.
Starburst libraries are computed for a variety
of stellar initial mass functions (IMFs). 
For all scenarios, the IMF is taken from \citep{kro93} after checking that  other classical IMFs
do not significantly change our predicted colours and template SEDs.
Star formation is initiated at the redshift of formation that we denote
as z$_{for}$\ taken as 10 when cosmic time slowly varies for redshift increases.
While this would nominally imply  an age of  13\,Gyrs  (adjusted to
lower redshift for Irregulars, 9 Gyrs) for local z=0 galaxies, this
should not be taken as the average age of the stellar population as
this is only when the star formation is initiated and not when the bulk
of the stellar population forms (this depends on many factors of the models). 
Changing $z_{for}$ from z=10 to z=20 or 30 will only vary typical
ages of local templates by a small amount ($\leq 0.4$\,Gyrs) and thus
resulting in only inconsequential variations in the SEDs.

\subsection{\bf Methodology} 
An automatic procedure  which uses a $\chi^2$\ minimization searches
for the best fit of observations. One or the sum of two spectral
templates are from the set of synthetic libraries built for starburst plus a variety
of SEDs fitting local Hubble type galaxies. For high redshift galaxies we work 
with template SEDs at redshift z in the observer's frame, by
using the z=0 template SEDs reproducing luminosities and colors of local galaxies. They are
corrected for cosmological expansion (redshift k$-$correction) and for evolution
(age e$-$correction) using the cosmic time$-$z relation.  For photometry
through various filters, apparent magnitudes $m^i_{\lambda}$\ of galaxy
type $i$\ through filter $\lambda$\, at age $t$\ and redshift $z$\ are computed by
\citep{rv88}:

\[
m^i_{\lambda}(t,z)= M^i_{\lambda}(t_0,0)+(m-M)_{\rm bol} + k^i_{\lambda}(z,t) + e^i_{\lambda}(z,t)
\]

M$^i_{\lambda}(t_0,0)$\ is the local absolute magnitude by type $i$,
$(m-M)_{\rm bol}$\, is the distance modulus and $k^i_{\lambda}(z,t)$\
and $e^i_{\lambda}(z,t)$\ are the expansion and evolution corrections,
computed from \pegase.3 synthetic templates.

\begin{figure*}
\centering
\includegraphics[angle=-90,width=8cm]{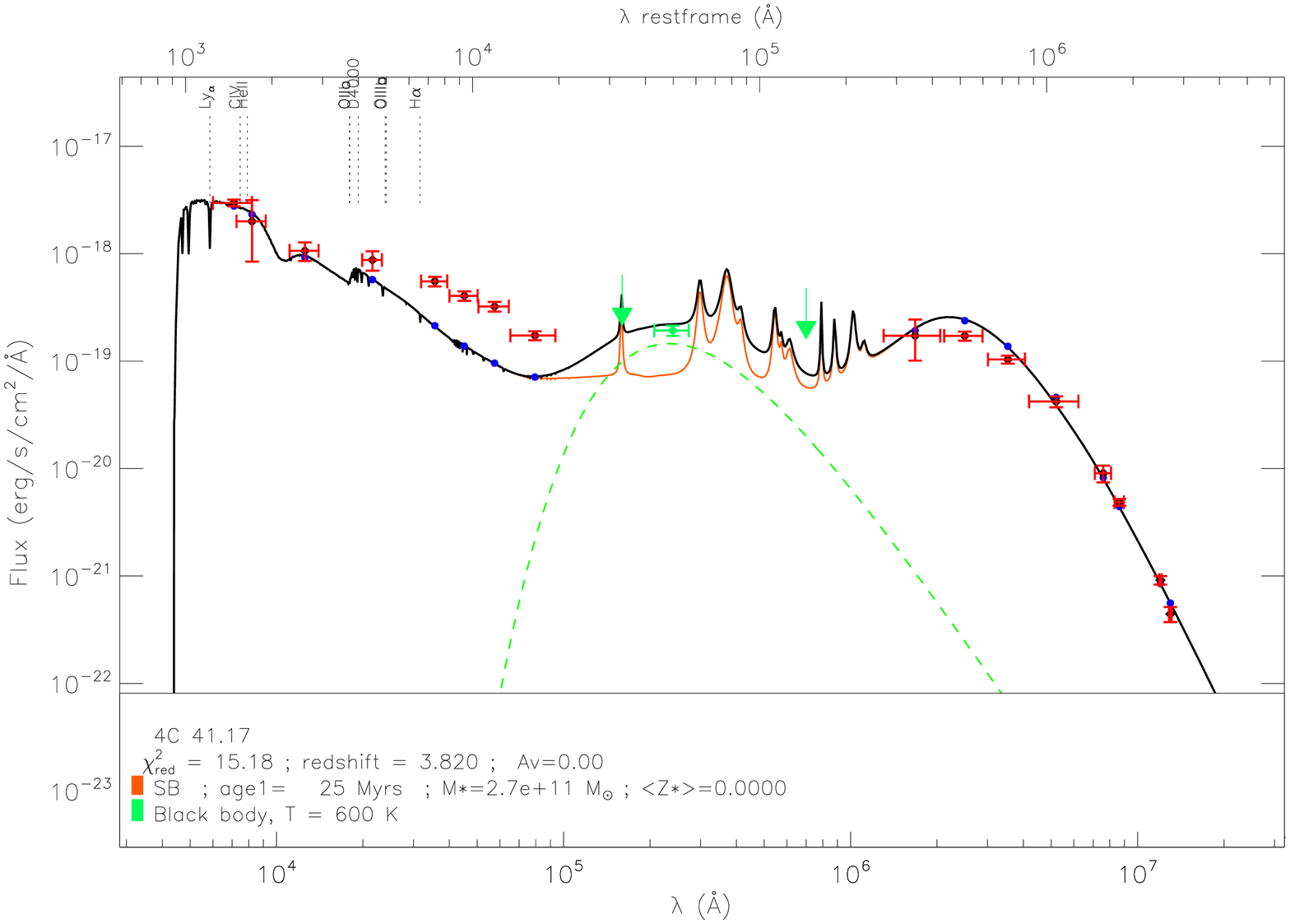}
\includegraphics[angle=-90,width=8cm]{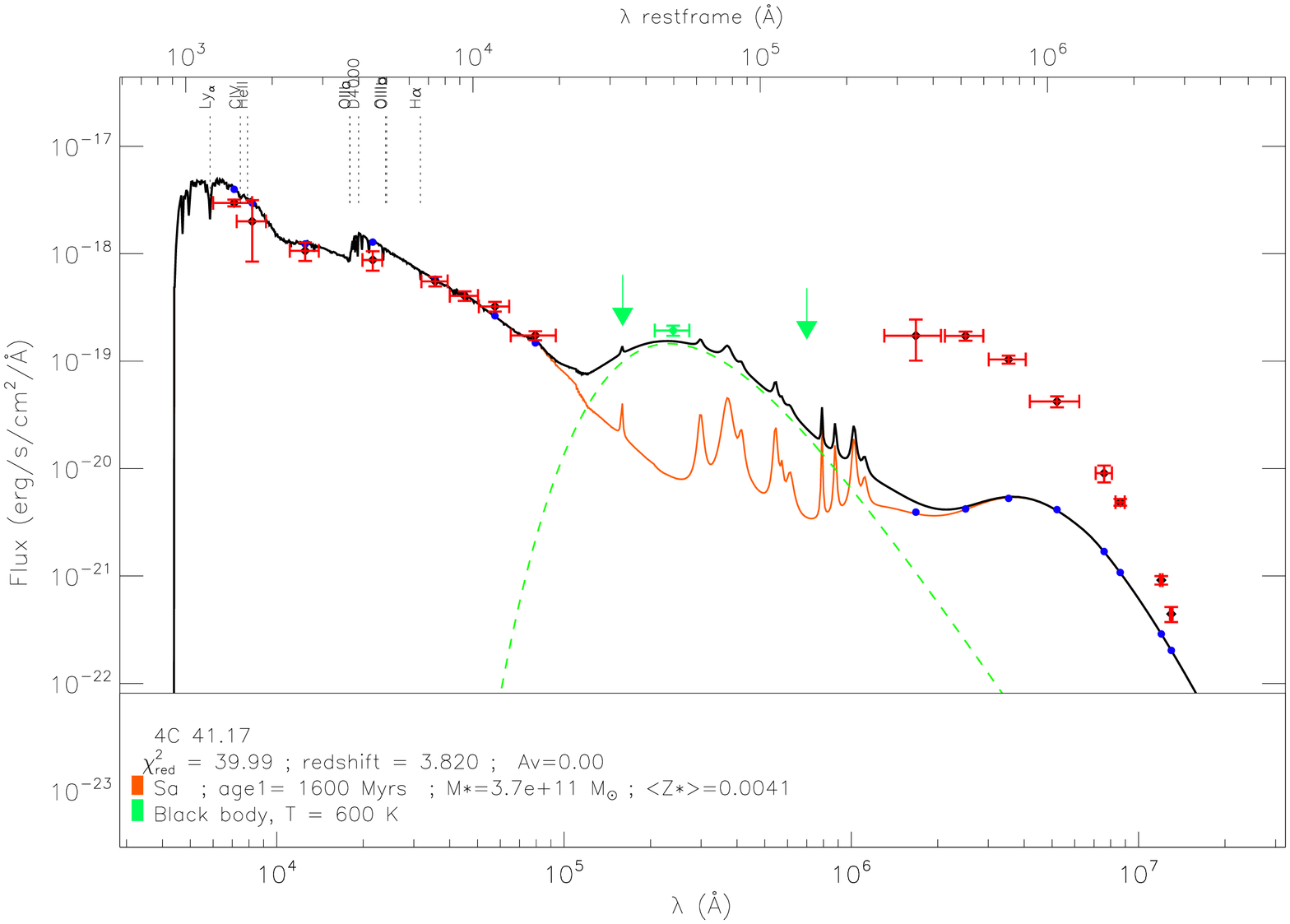}
\caption{The multi-wavelength SED of 4C~41.17 (dots with red error bars) compared
to the sum (black line) of two components: the AGN black body at
$T_{\rm eff}=600\,$K (dashed green line) plus a stellar component:
(left) a starburst at age 25 Myr; (right) a spiral Sa at age (time since
z$_{\rm for}$) of 1.6 Gyr.  The age, stellar metallicity, total stellar
mass are derived from the best fitting model.
Both of these single component models are missing a significant fraction
of the total energy, either in the near-IR for the starburst or in the
far-IR for the Sa-type galaxy evolutionary scenario.  Vertical dotted
lines mark the position of prominent emission lines.  A second component
is clearly needed to fit the SED. No intergalactic extinction ( $A_V=0$) is 
taken into account in models.}
\label{SaSB4C4117}
\end{figure*}

\begin{figure*}
  \centering
\includegraphics[angle=-90,width=16cm]{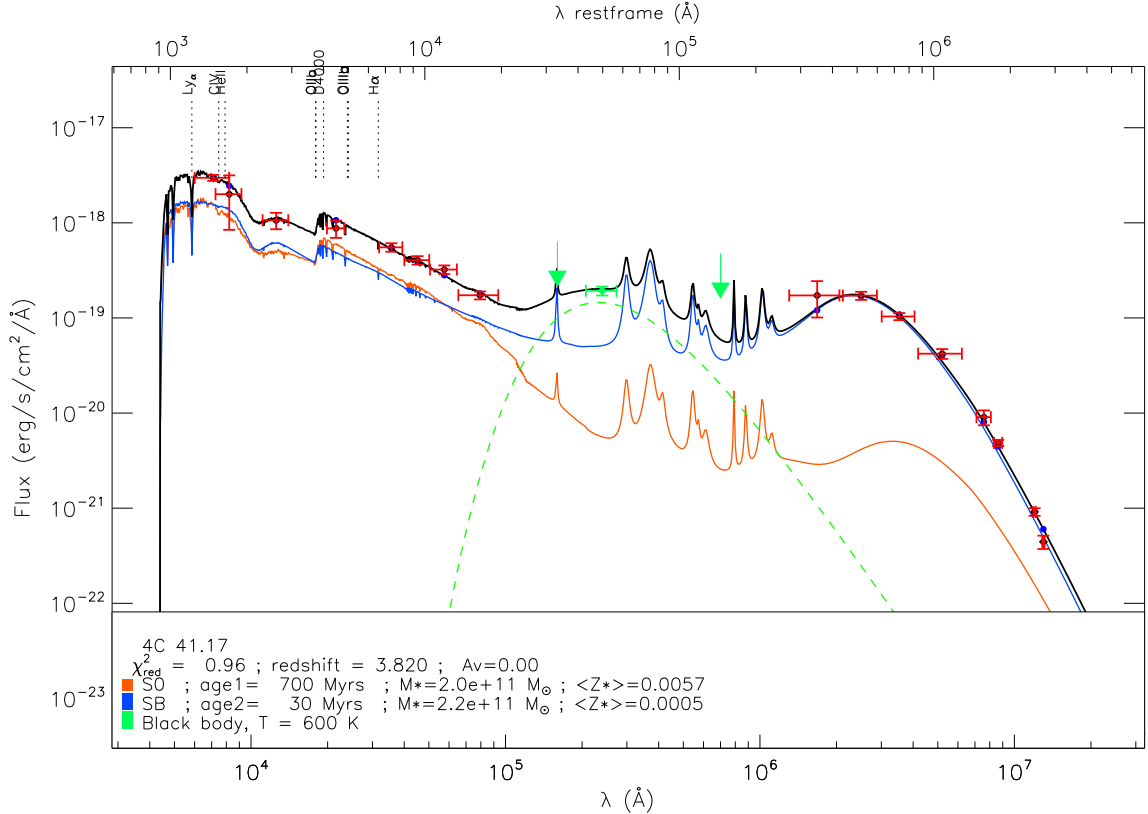}
\caption{The SED of 4C~41.17 (dots with red error bars) 
including the new {\it Herschel} data is best fitted (reduced
$\chi^2$\,= 0.96) by the sum (black line) of the S0-type population
evolution scenario at age (time since z$_{\rm for}$) of 0.7\,Gyr (orange
line) and a starburst with an age of 30\,Myrs after the initial 1 Myr
duration burst (blue line) from a dense ($10\times\,$standard NHI$_{\rm ISM}$)
medium.  A T$_{\rm eff}=600\,$K black body is used to model the thermal
emission from the AGN (green dashed line). Observations not included in
stellar fits are displayed in green. As can be seen, by excluding these data,
our results are relatively insensitive to the exact model adopted for the
thermal emission from the AGN.  Vertical dotted lines mark the position
of prominent emission lines. No intergalactic extinction ( $A_V=0$) is 
taken into account in models. }
\label{SEDs4C41.17}
\end{figure*}

\section{\bf  Results of the SED fitting }

Because the two $z$=3.8 radio-galaxies are selected for their small
AGN contribution to the UV/optical portion of their spectral energy
distributions as evidenced by their low level of polarization, simple
models of the thermal emission from the AGN are sufficient for this
analysis. The mid-IR SEDs provide an upper limit to the AGN component.
We modeled the AGN emission of 4C~41.17 with a black-body at $T=600\,$K
and of TN~J2007$-$1316 with the torus c model of \citep{pk92}
and \citep{kb88},see \citep{dro12} for details.
A more robust statistical analysis of the contribution of the AGN in
radio galaxies will be conducted for the whole of the HeRG\'E sample. For
the two radio galaxies studied here, the range of plausible AGN thermal
continuum models has only a minor impact on the results of our stellar
SED modelling and will not be discussed further.

\begin{figure*}
\centering
\includegraphics[angle=-90,width=14cm]{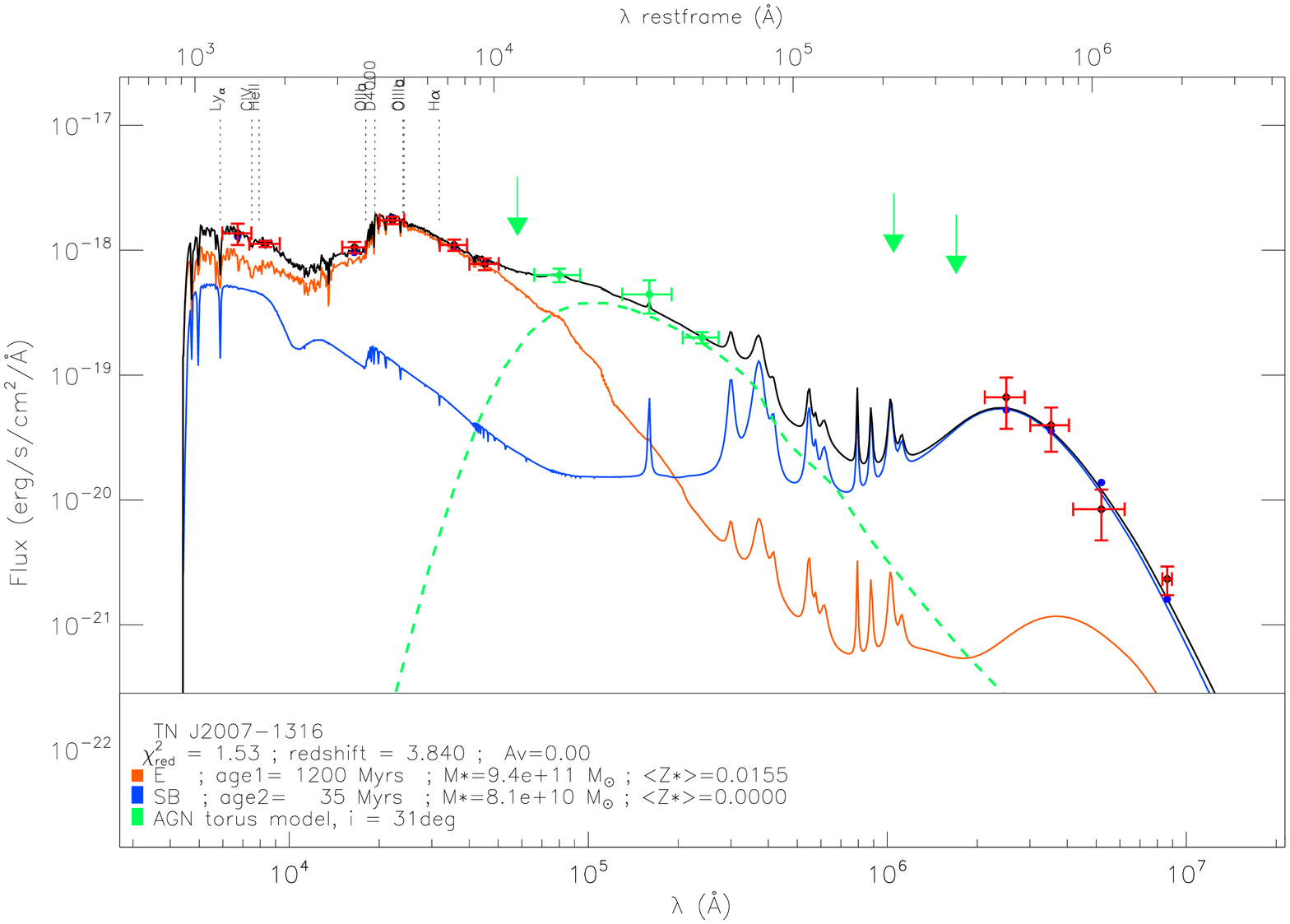}
\caption{The multi$-$wavelength SED of TN~J2007$-$1316 (dots with red error bars)
including the most recent Herschel data is fitted (reduced $\chi^2$=1.53) 
by the sum (black line) of an old (1.2\,Gyr) elliptical galaxy
(orange line) plus a starburst of 35\,Myrs (blue line)
from a dense medium of $10\times$\ standard NHI$_{\rm ISM}$ medium.  
The AGN model is the torus c model from Pier \& Krolik (1992). 
Data points shown in green are not included in stellar fits. 
Vertical dotted lines mark the locations of prominent
emission lines in the SED. No intergalactic extinction 
($A_V=0$) is taken into account in models.}
\label{SEDsTNJ2007}
\end{figure*}

\subsection{Single component model fits to the SEDs}

Fig.~\ref{SaSB4C4117}left shows the best fits of the observed 4C~41.17
SED with one single burst of star formation with a duration of 1 Myr
which is passively evolving for 25 Myr and also a single component
spiral-like evolutionary scenario with an age of 1.6\,Gyr.  The starburst
evolutionary scenario well fits the optical and far-IR ({\it Herschel})
data but fails to fit the region of the rest-frame 0.4-2$\mu$m (K$_s$
and IRAC/{\it Spitzer}) SED.  The optical peak is due to stellar
photospheric emission and the far-IR/sub-mm peak is due to a fraction
of the N$_{Lyc}$\,photons absorbed by cold grains within environments of
recent star formation.  While the starburst episode is consistent with a
substantial portion of the SED, \citep{sey12} for PKS 1138-262), having to let it evolve for $\sim$25
Myr suggests that the far-IR peak results from dust produced in
and heated by supernovae and AGB stars while the optical (UV-redshifted)
emission is due to the photospheres of recently formed mass stars.
The AGN component, modeled as a blackbody with a temperature of 600 K,
is compatible with data.  The starburst mass, consistent with the best
fit of the SED, is $2.7 \times10^{11}\,M_\odot$. In case of instantaneous 
starburst, the average stellar metallicity $< Z_*>$\ corresponds to 
the gas metallicity when stars were formed (null for a primitive 
cloud). For low values of $< Z_*>$\ the far-IR emission is dominated by the 
self gas$-$enrichment of the starburst, measured by the parameter 
$Z_{C+Si}$\ which measures abundances of carbon and Silicon. The important point is that from the K$_s$
and IRAC/{\it Spitzer} photometry, no matter what the exact starburst
parameters are adopted, the mid$-$IR emission is not well fit.
On the other hand, adopting a model with an older stellar population,
such as an early$-$type spiral Sa scenario with an age of $1.6$\,Gyrs and
having a stellar mass M$_*$\ of 3.7$\times10^{11}\,M_{\odot}$\ while
fitting the optical, NIR and {\it Spitzer}/IRAC data rather well, does
not fit the far$-$IR emission (Fig.~\ref{SaSB4C4117}right).  We note, because
it is important evidence for the existence of an older population in both
radio galaxies, that the SED of an older population shows a peak at
around $\sim$1$\mu$m which is clearly observed.  However, despite obvious
advantages to both models in fitting the SED, it is clear that neither
single component model provides even an adequate fit to the overall SED.

\subsection {\bf Two component model fits to the SEDs}

Because the one component fits are inadequate but each individually have
features which uniquely fit important  regions of the overall SEDs, it
seems  logical  to  attempt  two  component  fits  which  realize  the
advantages of  each single component model.   To this end,  we fit two
components, one representing a recent  burst of star formation and the
other, a passively evolving episode  of past star formation (of course
constrained by the  age of the universe at  z$\sim$3.8).  The best fit
scenario for  the radio galaxy 4C~41.17  (Fig.~\ref{SEDs4C41.17}) is a
sum of  the two stellar  components: a vigorous starburst  observed at
age of 30\,Myr from a cloud of stellar metallicity $<Z_*>=5\times10^{-4}$\, and
mass  M$_*=2.2 \times10^{11}\,M_\odot$  and a  second population  of a
massive   S0-type  scenario   with   a  short   e-folding  time   (see
Fig.~\ref{S0spectra})  typical of early-types,  observed at  0.7 \,Gyr
after   its   initial   formation   epoch,   with   a   stellar   mass
M$_*=2.0\times10^{11}\,M_\odot$\        and        a       metallicity
$<Z_*>=5.7\times10^{-3}$. Results  on stellar components are weakly
sensitive to  the adopted form  of the AGN  emission (T$_{eff}=600K$\,
black-body  model)  in  the  portions  of the  SED  which  are  likely
dominated by the thermal IR emission from the AGN.

The results  on the stellar components  derived from the  best fit for
TN~J2007$-$1316  (Fig.~\ref{SEDsTNJ2007})  are  generally  similar  to
those  for  4C~41.17,  namely there  is  the  need for  a  young
starburst and an  older stellar population similar to  that of a local
massive galaxy  evolved by a  redshift of 3.8.  Specifically,  the fit
requires a starburst of very low metallicity evolved for 35\,Myrs with
a total  stellar mass of M$_*=8.1\times10^{10}\,M_\odot$  and a second
component consistent with that of an evolved massive early-type galaxy
observed  at age of  1.2 Gyr  (after z$_{\rm for}$) with  the average
stellar metallicity ($<Z_*>=1.5\times10^{-2}$) and a significant stellar
mass ($M_{\star}=9.4\times10^{11}\,M_\odot$).

\begin{table*}
\caption{Characteristics of the young post-burst at $z$=3.8: age,
stellar mass $M_{\star}$, initial star formation rate, bolometric luminosity
L$_{bol}$, dust/bolometric luminosity ratio, carbon+silicon Z$_{C+Si}$
mass fraction, column density factor K (NHI)= K $\times$ NHI$_{\rm ISM}$),
and number of type-II SNe.}
\begin{tabular}{c c c c c c c c c c c}  
\hline\hline  
Post-burst & age  &M$_*$  & SFR$_{init}$   & L$_{bol}$  &L$_{dust}$/L$_{bol}$& Z$_{C+Si}$& K (NHI) & n$_{SNII}$ \\
    & Myrs &  10$^{11}$M$_{\odot}$ & 10$^{5}$M$_{\odot}$\, yr$^{-1}$ & 10$^{46}$\ erg.s$^{-1}$ & & & & 10$^{7}$Myr$^{-1}$\\
\hline     
in 4C~41.17   &  30  & 2.2  & 2.6  & 8.0  & 0.96 & 0.026 & 10 & 7.1  \\  
in TN~J2007$-$1316  &  35  & 0.8  & 0.1  & 0.3  & 0.97 & 0.025 & 10 & 0.3 \\
\hline\hline    
\end{tabular}
\label{PostStarburst}
\end{table*}
 
\begin{table*}
\caption{Characteristics of the old-star population at $z=3.8$: age, 
stellar mass M$_*$, current star formation rate, bolometric luminosity, 
dust/bolometric luminosity ratio, Carbon+Silicate dust mass fractions, SNIa number, SNII number.}  
\begin{tabular}{c c c c c c c c c c c  }  
\hline\hline  
Early-type  & age & M$_*$  &SFR$_{current}$& L$_{bol}$ &L$_{dust}$/L$_{bol}$&Z$_{C+Si}$ & n$_{SNIa}$ &n$_{SNII}$\\
    &Gyrs& 10$^{11}$\ M$_{\odot}$ &  M$_{\odot}$.yr$^{-1}$ & 10$^{45}$\ ergs.s$^{-1}$ &  & 
 & 10$^{5}$Myr$^{-1}$ & 10$^{5}$Myr$^{-1}$\\
\hline     
4C41.17  & 0.7  &  2.0  & 243 & 8.2&  0.522 &0.026 & 0.8 & 19.2\\  
TN J2007-1316 & 1.2  &  9.4  & 77  & 10.0&  0.096 &0.023 & 3.8 & 6.7\\
\hline\hline    
\end{tabular}
\label{OldPopulation}
\end{table*}

\subsubsection{A massive post-burst in distant radio galaxies}
For  the  two $z=3.8$\ radio  galaxies,  fitting simultaneously  the
optical  ($\simeq1500$\AA\ rest-frame) and  crucially  the far$-$IR  SED
provided by the {\it  Herschel}/submm data suggests that both galaxies
are underwent a massive starburst several tens of Myr ago ( a post-burst).  
The modeling  of  the far-IR  emission  depends on  the
metal-enrichment of Carbon and Silicon, $Z_{C+Si}$, while the observed
UV-optical   emission  is   dependent  on   the  amount   of  absorbed
photospheric emission intrinsic to  the young stellar population.  The
main  characteristics  of the  fitted  young  post-burst scenario  are
summarized in Table.~\ref{PostStarburst}. The best fit is a result
of having to balance the  far$-$IR and optical luminosities and spectral
shapes, measured by the reduced $\chi^2_{red}$\ minimum. A younger age
increases  the number  of Lyman  continuum photons  but  decreases the
metal-enrichment, favoring  the direct photospheric  emission relative
to dust emission. Older ages  have the opposite effect. Because of this
dependence,  the  starburst  mass  and  age are  well  constrained  and
unique. All other parameters of Table.~\ref{PostStarburst} 
($Z_{C+Si}$\ metallicity ,  supernovae number, high density value) 
are  not free parameters, but derived  from chemical evolution.  
We recall that the mean  metallicity of stars
$<Z_{\star}>$,  as  indicated in  the  legends,  only  traces the  gas
metallicity  when  stars were  formed.   In  case  of starbursts,  this
parameter is the initial   gas   metallicity and remains constant with age 
while gas metallicity evolves and is largely dominated at age 
of 30 Myrs by the   intrinsic
gas-enrichment  due   to  the  passive  evolution   of  the  starburst
scenario.

In the radio galaxy 4C~41.17 
(Fig.~\ref{SEDs4C41.17} and Table.~\ref{PostStarburst}) the new {\it Herschel}/submm observations
are well modelled by the cold grain emission of the highly massive
and metal-enriched ($Z_{C+Si}$=0.026) stellar environments. This
post-burst population was triggered 30\,Myrs earlier in a starburst
with a duration of 1~Myr (instantaneous) with a star formation rate
of $2.6\times10^{5}\,M_{\odot}$.yr$^{-1}$
(current stellar mass corrected for already died stars at age of  30\,Myrs). 
The rapid evolution of most
massive stars (a few $10^7\,$yrs) means that they have already evolved
to SN and AGB stars when the post-burst is observed, enriching the
interstellar medium. This implies that supernovae explosions 
(n$_{SNII}=7.1\times 10^{7}\,$\ by Myr) and AGB stars with circumstellar envelops are
dominant. A dense medium with the high column density 
(a factor $K=10\times$\ the column density measured  in the Galaxy interstellar medium NHI$_{ISM}$)
is required by the far-IR data, implying $10^{22-23}$atoms\,cm$^{-2}$. Such
column densities have been found in radio galaxies \citep{mu10}.
This post-burst in the far-IR dominates the total luminosity ($96\,\%$).
In the radio galaxy TN~J2007$-$1316 
(Fig.~\ref{SEDsTNJ2007} and Table.~\ref{PostStarburst}), the starburst age is  35\,Myrs 
with a mass of $8.1\times 10^{10}\,M_{\odot}$\ and an  initial metallicity of
zero, likely a primitive cloud. This suggests that most of the necessary
metal enrichment to explain the far-IR emission ($Z_{C+Si}$=0.025)
was due to self-enrichment of the young stellar population.

\subsubsection{The importance of the $1\,\mu$m peak}

Fig.~\ref{SEDs4C41.17} and Fig.~\ref{SEDsTNJ2007} show the typical
energy distribution of an old stellar population at wavelengths of the
$K-$band and through {\it Spitzer}/IRAC filters 
(from the 4000$\,$\AA\ discontinuity to the typical $1\,\mu$m peak in rest-frame). 
The evolution of early-type
galaxies are characterized by a short ($\simeq$1 Gyr) intense star formation
episodes. This
old population is seen in both targets perhaps suggesting it is a generic
feature of radio galaxies.  For both radio galaxies, this component has
an age that is a significant fraction of a Hubble time at z=3.8. We recall that we have
characterized the ages as the time from a formation redshift $z_{for}$, which for 4C~41.17 and
TN~J2007$-$1316 is roughly 0.7\,Gyr and 1.2~Gyr respectively. However,
acceptable fits allow for a range of age at least a factor of 2 around
these nominal values.  We also note that the bulk of the stars in this
old population formed more recently than the values listed.  This peak
in the SED arising from evolved giant stars only evolves slowly and has been
identified at all redshifts. This result supports our analysis of
the Hubble $K-$band diagram \citep{rv04}, predicting
already massive and old galaxies at $z\geq$ 4. 
Table.~\ref{OldPopulation}
presents the characteristics of the evolved old component. Stellar masses
are respectively 2$\times$ 10$^{11}$ M$_{\odot}$ for 4C~41.17 and $\simeq$
10$^{12}$\,M$_{\odot}$ for TN J2007$-$1316. These masses are higher
than that of the young starburst mass and perhaps suggest previous episodes
of merging at earlier epochs.  The contribution of the AGN to MIPS/{\it
Spitzer} data, modeled by a simple black-body law at T$_{eff}=600\
K$ as shown on Fig.~\ref{SEDs4C41.17} (dashed green line), and by a
more sophisticated torus c-model (adapted from \citep{pk92} by
 \citep{dro12}) for the radio galaxy TN~J2007$-$1316 
(Fig.~\ref{SEDsTNJ2007},dashed green line), does not change our conclusions: for these two radio 
galaxies, the AGN emission
is not significant in the near-infrared domain at $z=3.8$.

\begin{figure}
\includegraphics[angle=-90,width=8cm]{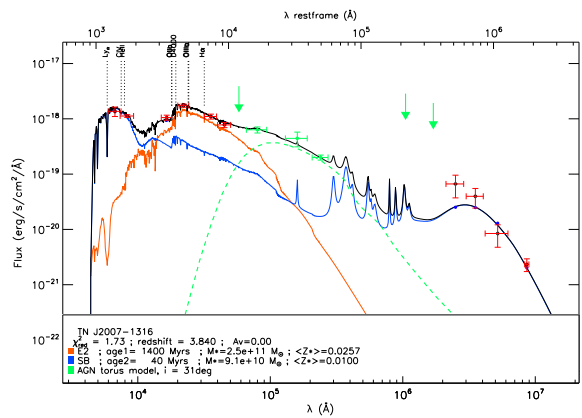}
\caption{Best fit (reduced $\chi^2$=0.98) of the multi$-$wavelength SED of 
TN~J2007$-$1316 (dots with red error bars) with the sum (black line) of a 40\,Myrs starburst
blue line) plus 
an early$-$type galaxy (1.4\,Gyr) after an episode of galactic winds 
(orange line) ejecting gas and dust, they stop any further star formation. 
 Masses and ages vary by $\leq50\%$\ compared to results of Fig.~\ref{SEDsTNJ2007}.} 
\label{Galwinds_TNJ2007}
\end{figure}

\section{\bf Discussion} 
The SEDs of the radiogalaxies 4C~41.17 and TN~J2007$-$1316
are revealed with two stellar, respectively young and old components and a simple AGN
model. The spectral synthesis used to fit the data is based on a variety
of template libraries including instantaneous starbursts and evolutionary scenarios
designed to fit the SEDs of local Hubble-type galaxies. Each scenario
is characterized by a star formation time-scale: $\sim$1 Myr for the
instantaneous starburst, $\sim$1 Gyr for ellipticals to $\sim$10 Gyrs
for spirals (see \citep{rv04}. In case of gas-dependent
SF laws, star formation time-scales are regulated by the accretion and
galaxy scale outflows of gas.  One template library is built for
each scenario. For starbursts, various standard IMFs
are modeled in these template libraries. Because in evolutionary models
as \pegase.3, IMFs are all normalized to $1\,$M$_{\odot}$ and is held fixed
throughout the evolution, changing the IMF does have an effect on relative
rate of metal-enrichment and dust fraction, but has only a minor effect
on the best fit results of the SED. Although only weakly constrained,
all of our best fits favor a \citep{kro93} initial mass function.
The most robust determinations from our fitting analysis are the
relative ages and total stellar mass of each best fit scenario.
For the two galaxies, our results suggest that shape of the SEDs
is due to a vigorous recent starburst superimposed on a relatively evolved
population whose formation was initiated at early epochs ($z_{for}\geq
10$; but this is not when the bulk of the stellar population was formed,
only when the star formation was initiated within the context of the
scenarios used to model the SEDs).  For the young starburst component,
our results suggest an age of $\sim$30 Myr with a stellar mass of $\simeq
10^{11}\,M_{\odot}$. The two peaks respectively seen in the optical and in the farIR/{\it
Herschel} are the data which constrain this fitted component the most.
This is because of the timescale for dust production in the postburst
AGB and SN populations.  Of much lower precision is the age of the older
component which is mainly constrained by the peak at rest-frame 1 $\mu$m
due to slowly evolving giant stars. This slow evolution means that a
wide variety of ages could in principle fit the data roughly equally
well, but all best fit models suggest relatively old ages (100s of
Myr). The best fits are that of an S0 galaxy with an age of 700\,Myr for
4C~41.17 and an elliptical galaxy scenario with an age of 1200\,Myrs for
TNJ~2007$-$1316. Both are early-types with a $\sim$1 Gyr time-scale and
masses of $\simeq  10^{11-12}\,M_{\odot}$.  
To check the sensitivity of the 4C~41.17 results to ages of the models,
we tested the relative variations in $\chi^2$\ for different models.
We maintain an instantaneous starburst as first component.  
Replacing the second component by another starburst 
of 1 Myr duration induces a $\chi^2$\
variation of ~10\% and an age reduction of 80\%
with deeper balmer lines. Replacing the second component by an 
later-type (Sa) scenario with star formation scale  of a few
Gyrs induces a $\chi^2$\ variation of  5\% and an age increase of ~120\%.

For TN~J2007$-$1316,
no other scenario, excepted ellipticals, is acceptable within a $\chi^2$
variation of 400\%. We also test (Fig.~\ref {Galwinds_TNJ2007}) a second
component of an early-type model suffering galactic winds at early epochs
(at age of 1\,Gyr). The 1$\mu$m peak is then better fitted to IRAC data
at age of 1.4\,Gyr and summed with a younger (40\, Gyrs) starburst. For
the two populations, ages do not vary more than 50\%.

In summary for these two powerful radio galaxies, the young post-burst and
an evolved component of a short time-scale ($\sim$1 Gyr) are the best
solutions. The timescale for both the increase in the metallicity with
time and relative fraction of dust in the ISM which produced the far-IR
emission are calculated in a self-consistent manner based on stellar
yields and the production of dust by stars.  The relatively poor time
resolution in fitting the broad band continuum with stellar synthesis
models for high redshift galaxies does not allow us to discuss anything
but simple evolutionary models (not for example successive short bursts
which would look more like continuous star formation in the models).
Long star-formation time-scales results in two cumulated effects:
increasing metal and dust fraction and simultaneously decreasing the
photospheric emission of massive stars.  Because of these effects, the
parameter-range which simultaneously fits both the rest-frame optical
and infrared portion of the SEDs is relatively narrow.  That is why
it is important to do this analysis on multiple sources, which if give
consistent results, means that the general characteristics of the best
fits are likely to be robust.

The general conclusions we can make are that the masses are likely in the
range 10$^{10-11}$M$_{\odot}$\ for recent starbursts and 10$^{11-12}$M$_{\odot}$\ for
the early-type evolutionary scenarios used to fit the data.  Starbursts
and evolved galaxies, have mass ratios of 1:1 to 1:10 typical of  major mergers,
so that any mass interchange will not significantly affect results.
Moreover all mergers are gas-rich inducing an efficient star formation.
\begin{table}
\caption{Starburst/old population ratios}
\begin{tabular}{c c c c c  }  
\hline\hline  
Radiogalaxy &  ages  & stellar mass & SNII number & factor K  \\ 
\hline     
4C41.17  &  30:700 &  2.2:2.0 & 71:1.9&  10:1 \\  
TNJ 2007-1316 &  35:1200 &  8:94  & 30:6.7&  10:1 \\
\hline\hline    
\end{tabular}
\label{ResultsMergers}
\end{table}

\subsection {Star formation process in 4C41.17}

What is triggering so intense star formation in distant radio galaxies?
One of the most striking feature of these distant SEDs is the double peak
of the starburst activity, respectively in the far$-$IR {\it Herschel} from
dust and the optical, emitted from star photospheres.  The instantaneous
starburst model evolves to the age of $\sim$30\,Myrs old likely dominated
by supernovae and AGB stars. The huge stellar mass 
$\simeq$\, 2.2\,10$ ^{11}$M$_{\odot}$\ corresponds to a rapid metal-enrichment from
a no-primitive cloud ($<Z_*>=5\,10 ^{-4}$).  The emission is issued from
a gas of high density, characterized by the factor $K$\ = 10, describing a ten time denser ISM
that the classical interstellar medium.
A model has been already proposed for 4C~41.17, a template of the
so$-$called alignment effect \citep{ch90,mi92}. Based on the jet-cloud 
interaction inducing star formation
by shocks, suggested by many authors from 1980, it was revisited by
\citep{bi00} in the light of HST and radio emission line data
\citep{vbg98,dey97}.  The authors conclude to a
set of parameters: baryonic mass of $\simeq$ 8$\times$10$ ^{10}$\,M$_{\odot}$,
a few Myrs star formation time-scale, a dynamic radio galaxy age of
$\simeq$ 30 Myrs and the high density of $\simeq$ 1-10 cm$^{-3}$. All
these values are compatible with our stellar population model, favoring
the jet cloud interaction model for triggering star formation at high
redshift.  Another favorable argument for this scenario, at a much lower
rate, is the star formation by jet-cloud interaction discovered in the
core of NGC 1068  with the high angular resolution of NaCo/{\it VLT}
\citep{ex11}.  

The dense medium measured by a column density factor $K$=10 meaning
10$^{22-23}$ atoms cm$^{-2}$\ only seen in the starburst component
requires justifications.  A better spatial resolution associated to
a more refined 3D spectroscopy with the new generation of instruments
(MUSE/VLT) will solve the location of star formation activity and the
feedback measurement.

The coincidence of the short timescales
of both starburst and jet-interaction could only be due to a similar
causal origin. In that case, the post-burst we are seeing at works is
only an episode of an active merger process. 
Many other arguments would favor massive gas$-$rich mergers at high
redshifts. The discovery of huge CO components \citep{deb05,eng10,ivi12} 
already suggest evidences of major mergers in such powerful
distant radio galaxies. Another convincing remark in favor of gas-rich mergers 
is that to form so huge stellar masses in short episodes requires large volumes
of high gas-density.   Table.~\ref{ResultsMergers} gives the characteristics of the
intervening masses. with mass ratios typical of major
mergers of young and evolved components of various metallicity,  
occasionally of primitive gas, implying several star generations through
merger processes.

\subsection {\bf The old early-type stellar population} 

Local early-type galaxies are generally not luminous in the far$-$IR
\citep{xi04}, which has been interpreted as the result of galaxy
winds driven by the evolved populations \citep{mb71}
The other possible mechanism for driving winds is the mechanical
and radiative energy from intense starbursts.  For example, the local
archetypal starburst galaxy M82, galactic winds were recently studied
with {\it Herschel}/PACS  \citep{cont12}. The authors propose
that cold clouds entrained in the galaxy disk outflow.  At high redshifts,
star formation activity is intense enough at the earliest epochs to induce
intense winds which may also entrain the molecular and neutral atomic
hydrogen driving it out of the galaxy. This assumption will possibly be
statistically analyzed on the totality of the HeRG\'E sample.

The new important result is the emergence in the mid-IR (optical
rest-frame) of an (already evolved) stellar component. Seen through the
specific window (IRAC filters).  The age and huge stellar mass of this old
population is calibrated on the SEDs of two selected radio galaxies, in
particular when galactic winds are imposed.  At age of $\sim$1\,Gyr, the
evolved stellar population is giant stars building the well-known peak at
1$\mu$m (rest-frame; Fig.~\ref{S0spectra}). All these results favor
the presence of highly massive early type galaxies initiated at early
epochs z$_{for}>$10, which must have grown very rapidly.  Because the
insufficient time resolution, the declining continuous star formation
laws, as used in \pegase.3 and other codes could also be considered 
as a series of successive starbursts with intensities declining with time.
From this point of view, which is also consistent with the data, the
hierarchical assembly history of most massive ($\sim$8$\times$10$^{12}$
M$_{\odot}$) haloes in a $\sim$3~Gpc$^3$\ volume as proposed in
\citep{li07} to form populations of z=6 quasars would be compatible
with our results.

\subsection{The link with radio-quiet massive galaxies} 

The discovery of these $z=3.8$ old stellar component in
powerful radio galaxies with a superposed starburst in a high density
medium, warrants a comparison with radio-quiet early-type high-$z$
galaxies. These radio quiet galaxies have median stellar masses 
of $\sim$4$\times$ 10$^{11}$
M$_{\odot}$ at z=2.0-2.7 \citep{kr06}. Their mass estimates are likely
underestimated because observations were through the
Gemini K$_s$\ filter, missing the population of dominant 
 rest-frame K-band emission redshifted in the
{\it Spitzer} filters. However the comparison with powerful radio galaxies 
is justified because their respective masses are of the same order.
 Moreover deep and high-resolution images obtained
with {\it HST}/NICMOS and /NIC2 and Keck \citep{vdo08,st08} 
are associated to the presence of massive disks of old stars
at high redshift of remarkably small sizes with a median effective
radius, $r_{eff}$=0.9kpc.  Cosmological surface brightness dimming
has a strong impact on the perceived morphology of distant galaxies
and likely explains generally why disks are basically undetectable at
z$\geq$ 1.5. The decrement in surface brightness due to cosmological
dimming is of $\simeq$ 7 magnitudes at redshift 3.8. Only the central disk
and spheroidal components are of sufficiently high surface brightness to
be detectable even in deep images \citep{st08}.

If radio-load galaxies are a subpopulation of the massive radio-quiet
galaxies which are undergoing an episode of vigorous star formation,
it may well be that the powerful AGN 
illuminate the surrounding gas
and dust in the disk over larger scales than would not otherwise be visible
due to surface brightness dimming. Even the continuum emission would be
affected due to scattering of the AGN continuum by dust.  Moreover, in the
population of radio-quiet massive galaxies, the stellar population does
not have the massive young stellar component we have observed in IR-luminous powerful 
radio galaxies,
and therefore would be less luminous in both the far-IR and in the optical
perhaps below the threshold of the detectability. The massive evolved starburst
in IR-luminous powerful radio galaxies 
contribute to the illumination of the disk over larger scales, sticking out of the
surface brightness threshold.
Observing extended
emission preferentially in radio galaxies confirms the coexistence of the
radio and starburst activity in the short star forming episode perhaps
initiated by gas-rich mergers. Of course, the jet would also have an
impact through its interaction with the surrounding gas, compressing and
heating it.  This shaping of the gas by the radio jet may also lead to
stronger extended emission in both the emission lines and continuum.  

More knowledge on kinematics and apparent morphology strongly modified by distance effects
are also required for justifying the spheroidal structure of early-type
galaxies.  In case of radio galaxies, all these properties have to be
associated to the presence of a supermassive black hole. Future studies
will focus on theluminosity  relationship between AGN and star formation activity
of a larger sample of HeRG\'E galaxies.

\section{\bf Conclusions}
From the \pegase.3 evolutionary spectral synthesis of two $z=3.8$\, radio
galaxies, the SED fitting  on a large wavelength coverage (UV/optical
to far-IR/submm), identifies  two distinct stellar components.
The originality and specificity of the analysis is to work in the
observer's frame by using robust local templates, corrected at high $z$\
for the Universe expansion (k-corrected) and evolution due to distance
(e-corrected),  following scenario librairies by types of the code \pegase.3.
At any time, synthetic templates are continuously computed with coherent
UV/optical stellar emission and dust absorption  reemitted in the
far-IR. Instantaneous (1Myr) starbursts and a variety of scenarios of
the Hubble sequence, varying with
1-10 Gyrs star formation time-scales, are considered for building template libraries. Best
SED fits are derived in the observer's frame by a $\chi^2$\ algorithmic procedure  on the sum
of two stellar components plus a simple AGN model.

Main results are similar for the two galaxies at $z=3.8$:
  
$\bullet$ one single component is unable to explain the complete spectral energy
distribution, whatever types and ages.

$\bullet$ the sum of two stellar components give the best SED fits, following:

$\bullet$ A 1 Myr starburst, observed at age of 30\,Myr, forming a huge
stellar mass $\simeq 10^{11}$ M$\odot$\ in a dense ($10 \times$ standard ISM) medium, dominant in the far-IR {\it
Herschel} to submillimetric and in the optical domains. 

$\bullet$ The discovery in the K-band to {\it Spitzer}/mid-IR  domains
(1$\mu$m rest frame) of an already evolved  ($\simeq$1\,Gyr at $z=4$) population. 

$\bullet$ The evolved stellar component is compatible with an early-type galaxy
( S0-Elliptical) initiated at early epochs (z$_{for}\geq 10$), with possible galactic winds.

$\bullet$  The mass of the evolved component is huge of 10$^{11-12}
M_\odot$\ for the two galaxies at $z\simeq$4.

$\bullet$  The possibility of a gas-rich merger (joined to a possible jet-cloud interaction) 
is favored.

Over the longer term, we intend to include a more refined modeling
of the AGN contribution to the overall SED in the \pegase.3 evolutionary
synthesis models and to model the optical, {\it Herschel}/submm and
{\it Spitzer}\ SED data of the HeRG\'E sample over its entire redshift range
($1 \leq z \leq 5$).

\section{\bf Acknowledgments}
This work is based on observations taken at the European Southern Observatory with the Very Large Telescope, Paranal, Chile, 
with program ID 069.B$-$0078. It is also based in part on observations made with Herschel, 
a European Space Agency Cornerstone Mission with significant participation by NASA.

We finally thank the referee Dr. Alan Stockton for his helpful comments and
suggestions which aided us in clarifying our arguments.


\end{document}